\documentclass[12pt]{preprint}
\usepackage[dvips]{graphicx}
\usepackage{xcolor}

\usepackage{bm}
\usepackage{amsfonts}
\usepackage{amssymb}\usepackage{times}
\usepackage{sub_JP}
\usepackage{MHequ}

\definecolor{carlos}{HTML}{5679EE}
\definecolor{jpe}{HTML}{FF0000}

\def\d{\mathrm{d}}
\def\kB{k_\mathrm{B}}
\def\bx{{\bm x}}

\def\bxi{{\bm \xi}}
\def\xt{{\bm x}_t}
\def\nablax{\partial_{\bm x}}
\def\v{{\bm v}}
\def\A{\mathcal{A}}
\def\C{\mathcal{C}}

\def\ri{\rho_{\rm i}}
\def\rf{\rho_{\rm f}}

\let\rho=\varrho
\def\eref#1{eq.~(\ref{#1})}
\def\Eref#1{Eq.~(\ref{#1})}
  \def\fref#1{Fig.~\ref{#1}}
\def\Label#1{}

\begin{document}

\title{Revisiting the Monge problem in the Landauer limit} 
\author{Jean-Pierre Eckmann${}^{1,2}$, Carlos~Mej\'ia-Monasterio${}^3$}
\institute{${}^1$D\'epartement de Physique Th\'eorique, Universit\'e de
Gen\`eve\\${}^2$Section de Math\'ematiques, Universit\'e de
Gen\`eve\\${}^3$School of Agricultural, Food and Biosystems
Engineering, Technical University of  Madrid}
\date{\today}
\maketitle

\begin{abstract}
  We discuss the Monge problem of mass transportation in the framework
  of stochastic thermodynamics and revisit the problem of the Landauer
  limit for finite-time thermodynamics, a problem that got the interest of
  Krzysztof Gawedzki in the last years. We show that restricted to one
  dimension, optimal transportation is efficiently solved numerically
  by well known methods from differential equations. We add a brief
  discussion about the relevance this has on optimising the processing
  in modern computers.
\end{abstract}

\section{Introduction}

The last decades have seen an increased interest in establishing a
thermodynamic understanding of nonequilibrium processes occurring in
small systems.  However, unlike macroscopic systems to which
thermodynamics apply, small systems are characterised by relatively
large fluctuations and time scales that are comparable to their
relaxation times, rendering a thermodynamic description inappropriate.

A great  advance in our  understanding of the thermodynamics  of small
systems  has  been  made  in  the  last  decades  due  mainly  to  two
developments \cite{Helsinki2013}: on the one hand the discovery of the
nonequilibrium  fluctuation  relations  \cite{ECM1993},  relating  the
statistics of  molecular fluctuations with the  microscopic symmetries
of the dynamics of small  systems far from equilibrium, and ultimately
giving  theoretical support  for the  validity  of the  second law  of
thermodynamics  \cite{ES94,GC95,Jarzynski1997,LS1999,Rondoni2007}, and
on the  other hand the  development of stochastic  thermodynamics that
derives  a thermodynamic  framework valid  for the  single fluctuating
trajectories  \cite{Sekimoto1998,Sekimotobook,Seifert2012,Pelitibook}.
These results have been successfully applied to a broad range of disciplines,
including        physics        \cite{Ciliberto2017},        chemistry
\cite{Schmiedl2007b},  biological  systems \cite{Seifert2012},  active
matter \cite{Speck2016},  and computer  processing \cite{Wolpert2019},
among a host of many others.

One such nonequilibrium process is life itself, maintained at the
molecular level by molecular motors determining the finely tuned
kinetics of the cell. Molecular motors are responsible of essential
life processes such as vesicle transport or cell division
\cite{Juelicher1999,Seifert2011}, and the synthesis of ATP \cite{ATPsynthase}.
Notwithstanding  the  strong  fluctuations these  molecular  complexes
display,  it is  of  central importance  to  understand the  efficiency
molecular motors operate.

Another instance in  which efficiency plays a central role  is that of
processing in  modern computers.  On  the one hand,  every computation
involves an energetic  thermodynamic cost. On the other  hand, due to
the miniaturisation of computer processors, modern transistors operate
at scales at which stochastic thermodynamics applies. However, scaling
of  computer transistors  presents  nowadays technological  challenges,
mostly  related to  the  increase of  dissipation,  thus limiting  the
scaling up  in the number  of transistors and therefore,  the possible
optimisation of the speed  of computation \cite{Crutchfield2022}. This
calls for the development of highly efficient computing processes that
operate at minimal dissipation.

A fundamental  process for  computers is that  of memory  erasure. The
minimal amount of energy required to erase a bit of memory is given by
the   celebrated  Landauer   limit  \cite{Landauer1961},   stating  an
energetic lower bound  of $\kB T \ln 2$, where  $T$ is the temperature
and $\kB$ the Boltzmann constant. This energy is eventually dissipated
into the environment as heat. The Landauer limit is achieved when
the erasure of a bit occurs along a quasi-static process. In practice,
memory swaps happen at finite short time scales in a process which is
inherently noisy and out of equilibrium. 

Fourteen years ago Krzysztof got interested on the mathematical
formalisation of (at that time incipient area), nonequilibrium
fluctuation relations \cite{Chetrite2007}, as well as
on the Landauer limit \cite{Aurell2012}.

The  memory   erasure  process  can   be  stated  as  a   finite  time
nonequilibrium process  between an  initial state at  which a  bit is
observed to be in a specific state  and a final state in which the bit
is absent, and  indeed the Landauer limit  was experimentally verified
like this  \cite{Berut2012}.  In general,  an external control  on the
nonequilibrium  transition can  be devise  to obtain  optimal processes
that minimise \emph{e.g.}, dissipation \cite{Schmiedl2007,Aurell2011}.
In particular,  in Ref.~\cite{Aurell2011} such optimisation  was shown
to be solved for the Langevin  dynamics in the overdamped limit by the
Monge-Kantorovich   optimal    mass   transport   and    the   Burgers
equation.  This is  particularly  relevant in  computer processing  as
there one searches for minimal dissipation  but at the same time, fast
processing.

In 2012, Krzysztof and his colleagues published a paper
\cite{Aurell2012} in which they related the Landauer Principle
\cite{Landauer1961} to the Monge-Kantorovich optimal mass
transport. In a talk by Krzysztof in Geneva in April 2013, ``2nd law
of thermodynamics for fast random processes'' he showed how the
Landauer Principle is related to an overdamped Langevin evolution from
an initial state to a different final stated. In discussions with him,
the question came up on how to efficiently and precisely compute the
dissipated energy. There were many methods around at the time, and
Krzysztof based his discussion on the papers by Benamou and Brenier
\cite{BB1,BB2}. He also gave a sequence of 3 talks in Helsinki on this
subject ``Fluctuations relations in stochastic thermodynamics'' where
an extensive list of references can be found
\cite{Helsinki2013}. Since the authors of \cite{Aurell2012} restricted
the discussion to 1 dimension, one can use methods from differential
equations which are more precise than the methods the used in that
paper. This will be part of our contribution, which illustrates the
breath of Krzysztof's interests and competence. We miss him.

\section{Stochastic Thermodynamics}
\label{sec:stochtherm}

We consider finite time nonequilibrium transitions  in $d$ dimensions,
with dynamics  described by  the Langevin  equation in  the overdamped
limit
\begin{equ} \label{eq:Langevin}
\d\xt  = -\mu \ \nablax U(\xt ,t) \d t + \d\bxi_t \ ,
\end{equ}
where $U(\xt ,t)$ is a smooth control potential and $\bxi_t$ white noise
with zero mean $\langle \d\bxi^i_t \rangle = 0$, and covariance $\langle \d\bxi^i_t \ \d\bxi^j_{t^\prime} \rangle = 2 D^{ij}
\delta(t-t^\prime) \d t$.

The diffusion matrix $D$ and  mobility matrix $\mu$ appearing above, are
assumed positive and satisfying the Einstein relation $ D = \kB T \mu$
where $\kB$  is the Boltzmann constant,  so that the noise  models the
fluctuations of a thermal bath at temperature $T$. 

During  the  nonequilibrium  transition  we assume  that  the  control
potential  changes   from  $U(\bx_0 ,0)   =  V_i(\bx)$   at  time   $t=0$  to
$U(\bx_\tau ,\tau) = V_f(\bx)$ at time $t=\tau$. Given an initial probability
density $\rho(\bx_0,0) = \rho_i(x)$, $\xt $ defines a Markov diffusion
process for times $t>0$, with generator
\begin{equ} \Label{eq:Lt}
  \mathcal{L}_t = -\left(\nablax U(\xt ,t) \right) \cdot \mu
  \nablax + \kB T \nablax \cdot \mu \nablax \ .
\end{equ}
The probability density evolves according to the Fokker-Planck
equation
\begin{equ} \label{eq:F-P}
  \partial_t \rho(\xt,t) = \mathcal{L}_t^\dagger \rho(\xt,t) \ ,
\end{equ}
where 
\begin{equ} \Label{eq:adjoint}
  \mathcal{L}_t^\dagger = \nablax \mu \left(\nablax U(\xt ,t) \right) \cdot \mu
  \nablax + \kB T \nablax \cdot \mu \nablax \ ,
\end{equ}
is the adjoint of $\mathcal{L}_t$.

Following  Ref.~\cite{Sekimoto1998}, the  energy balance  for the  single
fluctuating trajectories of the process \eref{eq:Langevin} leads to the
framework   of  stochastic   thermodynamics,  developed   to  give   a
thermodynamic description to small systems in contact with a heat bath
and driven out of equilibrium (see \emph{e.g.}, Refs.~\cite{Seifert2012,Pelitibook}
for a modern review).

Defining the work done on the system during the time interval $\tau$
as
\begin{equ} \Label{eq:work}
  W = \int_0^\tau \partial_t U(\xt ,t) \d t \ ,
\end{equ}
and  the heat  released  by the  system into  the  environment in  the
Stratonovich convention as
\begin{equ} \Label{eq:heat}
  Q = - \int_0^\tau \nablax U(\xt ,t) \circ \d \xt \ ,
\end{equ}
the balance
\begin{equ} \Label{eq:1stlaw}
  W - Q = \Delta U \ ,
\end{equ}
with $\Delta U = U(\bx_\tau ,\tau) - U(\bx_0 ,0)$, expresses the
conservation of energy that holds for every fluctuating trajectory of
the transition process, in analogy to the First Law of thermodynamics.

To obtain a  fluctuating version of the Second  Law of thermodynamics,
we first notice  that the entropy change associated  with the transition
from time $0$ to $\tau$, can be split into two contributions
\begin{equ} \label{eq:S}
\Delta S_{\mathrm{tot}} = \Delta S_{\mathrm{sys}} + \Delta
S_{\mathrm{env}} \ .
\end{equ}
The first term on the right hand side of \eref{eq:S} corresponds to
the entropy change of the system due to the evolution of the probability
density
\begin{equ} \label{eq:DSsys}
\Delta S_{\mathrm{sys}} = S_{\mathrm{sys}}(\tau) - S_{\mathrm{sys}}(0) \ ,
\end{equ}
where
\begin{equ} \Label{eq:Ssys}
S_{\mathrm{sys}}(t) = -\kB \int \rho(\xt ,t) \ln
\left(\rho(\xt ,t)\right) \d \xt  \ ,
\end{equ}
is simply the Gibbs-Shannon entropy  with respect to the instantaneous
probability density.

The second contribution  on the right hand side of \eref{eq:S}
corresponds to the change of entropy of the environment due to the
dissipated heat
\begin{equ} \label{eq:DSenv}
  \Delta S_{\mathrm{env}} = \frac{1}{T} \langle Q \rangle \ ,
\end{equ}
where $\langle Q \rangle$ is the mean heat released during the
transition.

To obtain the second law it is expedient to define the current
velocity of the process $\v(\xt ,t)$ \cite{Nelsonbook}. We first note
that the instantaneous probability density for a Markov diffusion
process can be written as
\begin{equ} \Label{eq:rho}
\rho(\xt ,t) = \langle \delta\left(\bx - \xt \right) \rangle \equiv
\exp\left(-\frac{R(\xt ,t)}{\kB T} \right) \ .
\end{equ}
Moreover, the Fokker-Planck equation \eref{eq:F-P}, yielding the evolution of
$\rho(\xt ,t)$ can be rewritten as the advection equation
\begin{equ} \label{eq:adv}
\partial_t \rho + \nablax (\rho \v) = 0 \ ,
\end{equ}
in the current velocity defined as
\begin{eqnarray} \label{eq:v}
\v(\xt ,t) & = & -\mu\left( \nablax U (\xt ,t) + \frac{\kB T}{\rho(\xt ,t)} \nablax
                \rho (\xt ,t) \right) \nonumber \\
  & = & -\mu \ \nablax \left(U (\xt ,t)-R (\xt ,t)\right) \ .
\end{eqnarray}
The  current   velocity,  defined  through  an   appropriate  limiting
procedure     \cite{Nelsonbook,Aurell2012b,Aurell2012},    has     the
interpretation  of  the  mean  local  velocity  of  the  process  $\xt
$.   Correspondingly,  in   terms   of  the   current  velocity,   the
Fokker-Planck equation is equivalent to deterministic mass transport.

Now,  using  \eref{eq:adv},  the   entropy  change  of  the  system
\eref{eq:DSsys}, can be written after integration by parts as
\begin{eqnarray} \Label{eq:DSsys2}
\Delta S_{\mathrm{sys}} & = & \int_0^\tau \d t \dot{S}_{\mathrm{sys}}
  \nonumber \\
  & = & \int_0^\tau \d t \left(-\kB \int \left(1 + \ln(\rho(\xt ,t))\right)
        \partial_t \rho(\xt ,t)) \d \bx \right) \nonumber \\
  & = & \frac{1}{T} \int_0^\tau \d t \int \nablax R(\xt ,t) \cdot
        \v(\xt ,t) \ \ \rho(\xt ,t) \ \d \bx \ ,
\end{eqnarray}
where $\dot{S}_{\mathrm{sys}}$ is the time derivative of $S_{\mathrm{sys}}$.

In the same way, the change of entropy of the environment becomes
\begin{eqnarray} \Label{eq:DSenv2}
\Delta S_{\mathrm{env}} & = & -\frac{1}{T} \int_0^\tau \int U(\xt ,t)
                              \partial_t \rho(\xt ,t) \d \bx \nonumber \\
& = & -\frac{1}{T} \int_0^\tau \d t \int \nablax U(\xt ,t) \cdot
        \v(\xt ,t) \ \ \rho(\xt ,t) \ \d \bx \ .
\end{eqnarray}

Combining both contributions and using \eref{eq:v}, the total
entropy change \eref{eq:S} becomes
\begin{equ} \label{eq:DS}
\Delta S_{\mathrm{tot}} = \frac{1}{T} \int_0^\tau \d t \int
\v(\xt ,t) \cdot \mu \ \v(\xt ,t) \ \rho(\xt ,t) \ \d \bx \ .
\end{equ}
This expression implies immediately the fluctuating analogue of the
second law of thermodynamics, namely
\begin{equ} \label{eq:2ndlaw}
\Delta S_{\mathrm{tot}} > 0 \ .
\end{equ}

\section{The optimal mass transport}
\Label{sec:Monge}

The mass  transport problem was  first considered by Gaspard  Monge in
1781  \cite{Monge1781}.  Roughly  speaking,  the  problem consists  in
calculating the most  economic way of moving a volume  of mass between
two places.  The modern approach  of Monge's optimal mass  problem was
formalised  by Kantorovich  in 1942  \cite{Kantorovich1942} (see  also
\cite{BB2,villani}). Optimal  mass transport  is nowadays  referred as
the Monge-Kantorovich problem. Here we adopt Monge's exposition.


Let $\rho_i(x)$ and $\rho_f(x)$  be two probability densities, bounded
and with compact support in the reals, and satisfying
\begin{equ} \label{eq:normcons}
  \int \rho_i(x) \d x = \int \rho_f(x) \d x = 1 \ .
\end{equ}

The optimisation problem is to find an
invertible smooth map $\varphi := \bx_f(\bx_i)$, that is measure
preserving, namely
\begin{equ} \Label{eq:transport}
  \int_{\bx_i(A)} \rho_i(\bx) \ \d\bx = \int_A \rho_f(\bx) \ \d\bx \ ,
\end{equ}
and minimises the objective function
\begin{equ} \label{eq:mincost}
  \int \C\left(\bx,\bx_f(\bx)\right) \rho_i(\bx) \ \d\bx \ ,
\end{equ}
where $\C\left(\bx,\bx_f(\bx)\right)$ is the cost transporting the
unit mass from its initial distribution $\rho_i$ into a final
distribution $\rho_f$. In its original formulation, Monge considered
the Euclidean distance as the cost function
$\C\left(\bx,\bx_f(\bx)\right)=\left|\bx -
  \bx_f(\bx)\right|$. However, the cost function can be taken as $\C\left(\bx,\bx_f(\bx)\right)=\left|\bx -
  \bx_f(\bx)\right|^r$. We will show that the case $r=2$ is
particularly relevant to formulate a refined Landauer limit for finite
time processes. This case was solved by Benamou and Brenier in 1999
\cite{BB1}, and was used by Krzysztof and colleagues in 2012 in
relation to the Landauer limit \cite{Aurell2012}.

\section{Minimal dissipation memory erasure}
\Label{sec:Landauer}

The  second  law  states  that for  any  thermodynamic  transformation
between two given  initial and final states, the  total entropy change
must be positive,  such as in \eref{eq:2ndlaw}.  This  is valid for
any  thermodynamic  transformation,  even  for  quasistatic  processes
occurring infinitely slowly.

Consider now a thermodynamic transformation constrained to be
completed in a fixed finite time $\tau$.  Dissipation is naturally
expected to be larger than in the quasistatic transformation, and the
question that arises is: what is the minimal possible dissipation
produced in a finite-time transformation? This question was answered
in Ref.~\cite{Aurell2012} for isothermal stochastic systems
with Langevin dynamics in the overdamped limit of section
~\ref{sec:stochtherm}. 

This    question     is    particularly    relevant     to    computer
processing. Landauer cost of  information processing, stating that the
erase of a bit  of information is performed at a cost  of no less than
$\kB T \ln 2$ dissipated  heat \cite{Landauer1961}. The Landauer limit
continues to be  the main reference in  information processing because
the process of  bit erasure is the elementary  operation that produces
maximal dissipation in universal computing with transistor logic gates
\cite{Crutchfield2022}.

In this section  we review briefly this optimal  solution by following
Ref.~\cite{Aurell2012}.

Consider the stochastic process $\xt$ of \eref{eq:Langevin}, driven
out  of   equilibrium  by   the  control   $U(\xt,t)$  from   a  state
$\rho_i(\bx)$  at  time  $t=0$  to   a  state  $\rho_f(\bx)$  at  time
$t=\tau$. The  goal is  to obtain  the optimal  choice of  the control
$U(\xt,t)$  minimising  the  dissipation,  as  given  by  \eref{eq:DS},
required to drive  the system along such transformation,  over all the
densities $\rho(\xt,t)$ and all velocity fields $\v(\xt,t)$ satisfying
\eref{eq:adv}, under the constraint
\begin{equ} \label{eq:const}
  \rho(\bx_0,0) = \rho_i(\bx) \ , \qquad \rho(\bx_\tau,\tau) =
  \rho_f(\bx) \ .
\end{equ}
In other words, we need to minimise the functional
\begin{equ} \label{eq:A}
\A\left[\rho_i,\v\right] = \int_0^\tau \d t \int
\v(\xt ,t) \cdot \mu \ \v(\xt ,t) \ \rho(\xt ,t) \ \d \bx \ .
\end{equ}

Apart from  a factor $\tau$,  the minimisation of \eref{eq:A}  over the
fields $\rho(\xt,t)$ and $\v(\xt,t)$ subject to \eref{eq:adv} and
\eref{eq:const},  was solved  by Benamou and Brenier  \cite{BB1} (see
also \cite{Aurell2012}). There it was shown that the optimal velocity
current minimising \eref{eq:A} is gradient
\begin{equ} \label{eq:grad}
  \v(\xt,t) = \nabla \phi(\xt,t) \ ,
\end{equ}
where $\phi(\xt,t)$ is convex and a solution of the Hamilton-Jacobi
equation. \Eref{eq:grad} implies that the optimal solution
corresponds, through \eref{eq:v}, to the optimal control
$U(\xt,t)$, and that $\v$ is also the local velocity of the optimal
control. 
  
Restricting ourselves to  smooth velocity fields $\v$ such that the Lagrangian
trajectories $\bx(t)$ satisfy $\dot{\bx}(t) = \v(\bx(t),t)$, the
solution to the advection equation \eref{eq:adv} is given by
\begin{equ} \label{eq:rhosol}
  \rho(\xt,t) = \int \delta\left(\xt - \bx(t;\bx_i)\right)
  \rho_i(\bx_i) \ \d\bx_i \ ,
\end{equ}
where $\bx(t;\bx_i)$ denotes the Lagrangian trajectory that at time
$t=0$ passes through $\bx_i$. Under the controlled transformation, the
Lagrangian map $\bx_i \mapsto \bx_f(\bx_i)$ should transport the
initial density $\rho_i$ into the final density $\rho_f$. 

Substituting \eref{eq:rhosol} into \eref{eq:A}, we can replace the
minimisation of the functional $\A$ over the velocity fields $\v$ into
the minimisation of
\begin{equ} \Label{eq:Ax}
\A\left[\rho_i,\v\right] = \int_0^\tau \d t \int
\dot{\bx}(\bx_i ,t) \cdot \mu \ \dot{\bx}(\bx_i ,t) \ \rho(\bx_i) \ \d \bx_i \ .
\end{equ}
over the Lagrangian flows satisfying $\bx_i \mapsto \bx(\bx_i,\tau)
\equiv \bx_f(\bx_i)$ such that
\begin{equ} \Label{eq:const-1}
  \rho_f(\bx) = \int \delta\left(\bx - \bx_f(\bx_i)\right)
  \rho_i(\bx_i) \ \d \bx_i \ .
\end{equ}
This constraint is equivalent to
\begin{equ} \label{eq:const-2}
\rho_f\left(\bx_f(\bx_i)\right) \
\frac{\partial(\bx_f(\bx_i))}{\partial(\bx_i)} = \rho_i(\bx_i) \ ,
\end{equ}
where $\frac{\partial(\bx_f(\bx_i))}{\partial(\bx_i)}$ is the Jacobean
of the Lagrangian map. We require the Lagrangian map to be smooth and invertible,
with a smooth inverse $\bx_f \mapsto \bx_i(\bx_f)$.

Minimising first over time under the above constraints we realise
 that for a positive definite matrix $\mu$ the minimal Lagrangian
 trajectories correspond to straight lines
\begin{equ} \label{eq:minL}
   \bx(t;\bx_i) =  \frac{\tau-t}{\tau} \bx_i + \frac{t}{\tau}
   \bx_f(\bx_i) \ .
\end{equ}
Therefore, the optimal solution is completed once the functional
\begin{equ} \label{eq:cost}
  \C\left(\bx_f(\bx_i)\right) = \int \left(\bx_f(\bx_i) - \bx_i\right)
  \cdot \mu^{-1} \left(\bx_f(\bx_i) - \bx_i\right) \rho_i(\bx_i)
  \d\bx_i \ ,
\end{equ}
is minimised over all Lagrangian maps $\bx_i \mapsto \bx_f(\bx_i)$.

\Eref{eq:cost} corresponds to  the Monge-Kantorovich transportation
problem  of \eref{eq:mincost},  with  a quadratic  cost, solved  in
Ref.~\cite{BB1}, and in Refs.~\cite{Aurell2011,Aurell2012b,Aurell2012}
in the context of stochastic thermodynamics. In Ref.~\cite{Aurell2011}
it was shown  that minimisation of \eref{eq:cost} is  solved by the
Burgers  equation  over  the   velocity  potential  $\phi$,  and  mass
transport by the Burgers velocity field \eref{eq:adv}.


Once the minimiser $\bx_f(\bx_i) $ is obtained, the minimal value of
the functional \eref{eq:A} is
\begin{equ} \Label{eq:Amin}
  \A_{\mathrm{min}} = \frac{1}{\tau}  \C_{\mathrm{min}} \ ,
\end{equ}
where  $\C_{\mathrm{min}}$   is  the  value  of   the  quadratic  cost
\eref{eq:cost} over the minimiser Lagrangian map \eref{eq:minL}. Then it
follows that the  minimum entropy production during  a transition from
$\rho_i(\bx)$ to $\rho_f(\bx)$ satisfying \eref{eq:normcons} in a fixed
time $\tau$ is
\begin{equ} \label{eq:Smin}
  \langle \Delta S_{\mathrm{tot}} \rangle_{\mathrm{min}} =
  \frac{1}{\tau T} \C_{\mathrm{min}} > 0 \ .
\end{equ}

Finally,  \eref{eq:DSenv}  and  the  value  \eref{eq:Smin}  yield  a
Landauer bound for the average dissipated heat during the erasure of
one bit of information in overdamped Langevin dynamics
\begin{equ}  \Label{eq:Landauer}
  \langle Q \rangle \ge \frac{1}{\tau}  \C_{\mathrm{min}} + \kB T \ln
  2 \ .
\end{equ}

\section{Numerical solution of the assignation problem}
\Label{sec:num}

Given an initial $\rho_i(\bx)$ and final $\rho_f(\bx)$ densities, in
this section we deal with the problem of solving numerically the
assignation problem to find the minimiser of the Lagrangian map.  This
was done in Ref.~\cite{Aurell2012} by means of several methods. Direct
integration of the constraint \eref{eq:const-2} was found to become
unstable at values of $\bx_i$ for which the derivative
$\d\bx_f(\bx_i)/\d\bx_i$ diverges (and similarly for the inverse
map). Similar problems were also found using a rearrangement
``auction'' algorithm \cite{Brenier2003}.  After discussing this with
Krzysztof in Geneva in 2012, we decided to use another, hopefully
faster and more precise method. In the rest of this section we show
its implementation and discuss its performance and general
limitations.

We assume that  $\rho_i(\bx)$ and $\rho_f(\bx)$ are  given and without
loss of  generality, satisfy \eref{eq:normcons}.  The  minimiser of
the      Lagrangian     map,      namely      the     optimal      map
$\bx_i \mapsto  \bx_f(\bx_i)$ that  transports $\rho_i$  into $\rho_f$
and minimises the cost \eref{eq:cost}, satisfies \eref{eq:const-2}.

\begin{figure}[t]
\begin{center}
\includegraphics*[width=0.8\textwidth]{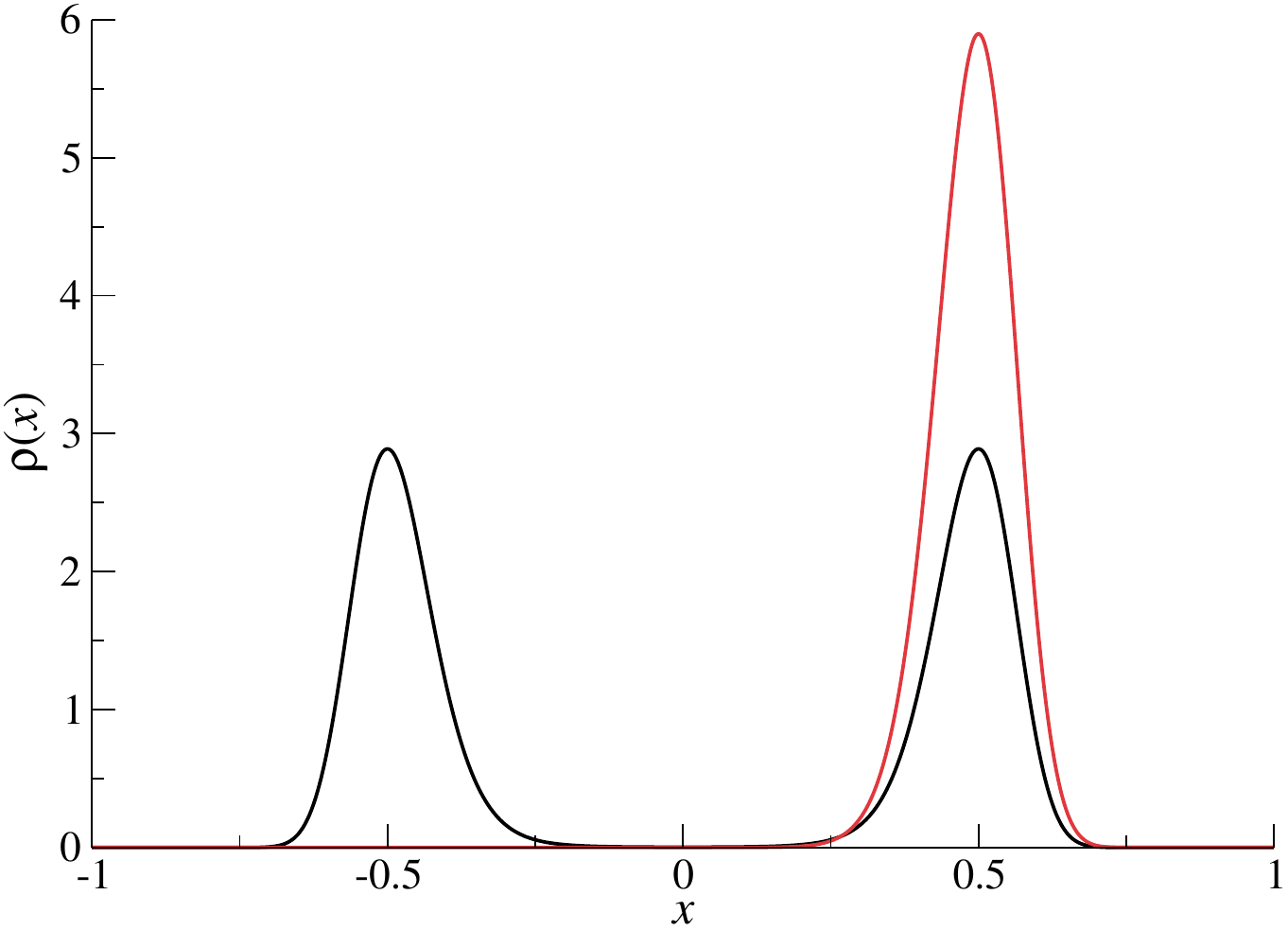}
\caption{The initial distribution $\rho_i(x)$ (black) and the final
  distribution $\rho_f(x)$ (red). While the problem is of course quite
  easy, intuitively, the issue is how to find the ``best'' numerical solution.
\label{fig:rhoirhof}}
\end{center}
\end{figure}

We   consider    the   optimal mass transport that    was   considered   in
Ref.~\cite{Aurell2012}:
\begin{equa}[eq:numeric]
  \rho_i(x)&=\frac{1}{Z_i}\exp\left(-\frac{a}{\kB T} \left(x^2-\alpha^2\right)^2\right)~,  \\
  \rho_f(x)&=\frac{1}{Z_f}\exp\left(-\frac{a}{\kB T} (x-\alpha)^2 \left((x-\alpha)^2+3 \alpha (x-\alpha)+4 \alpha^2)\right)\right)~,
\end{equa}    
where   $Z_i$   and   $Z_f$   are  normalisation   factors   so   that
\eref{eq:normcons}    is     satisfied,    and     the    constants
$a=112\kB T \mu\mathrm{m}^{-4}$ and  $\alpha =0.5 \mu\mathrm{m}$ were
chosen     to     match     the    experimental     realisation     of
Ref.~\cite{Berut2012}.  As  a  matter of  fact,  \eref{eq:Langevin}
models   the   dynamics   in    the   experiment   with   a   mobility
$\mu               =              \frac{0.213877}{\kB               T}
\frac{\mu\mathrm{m}^{2}}{s}$.   Furthermore,  the   control  potential
$U(\xt,t)$       is      effectively       one-dimensional\footnote{In
  \cite{Berut2012},  the experimental  potential along  the other  two
  dimensions is simply confining  and time independent.}.  The initial
and    final   densities    of   \eref{eq:numeric}    are   shown    in
\fref{fig:rhoirhof}.

The  minimiser  of   the  Lagrangian  map,  namely   the  optimal  map
$\bx_i \mapsto  \bx_f(\bx_i)$ that  transports $\rho_i$  into $\rho_f$
and minimises the cost \eref{eq:cost}, satisfies
\begin{equ} \label{eq:PDF}
  \int_{-\infty}^{\bx_i} \rho_i(\bx) \ \d\bx =
  \int_{-\infty}^{\bx_f(\bx_i)} \rho_f(\bx) \ \d\bx \ ,
\end{equ}
for all $\bx_i$, and uniqueness is guaranteed if for each $\bx_i$ one
chooses the minimal $\bx_f(\bx_i)$.

\begin{figure}[t!]
\begin{center}
\includegraphics*[width=0.8\textwidth]{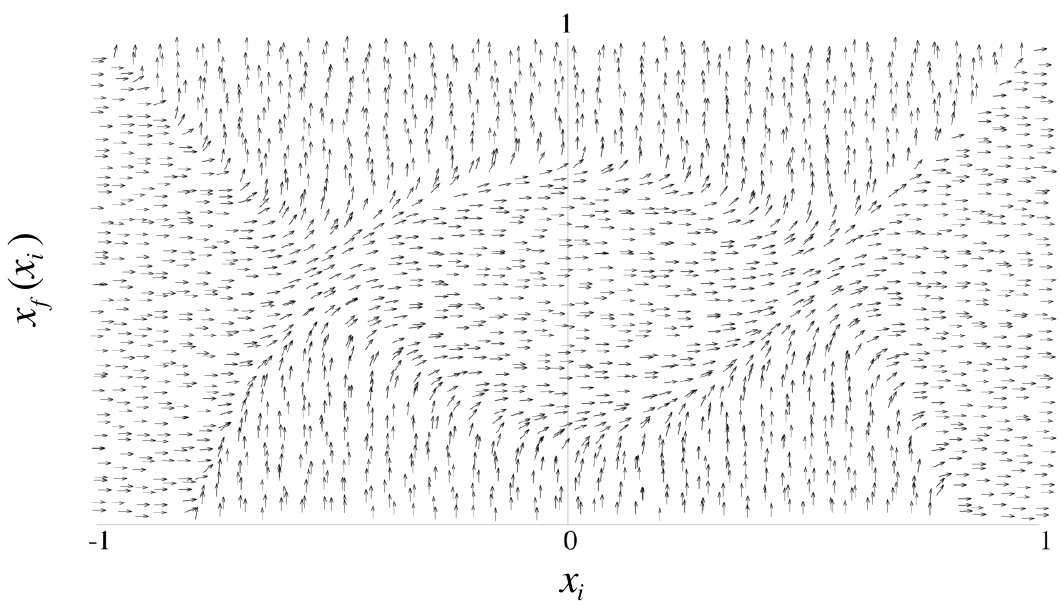}
\caption{Vector field of the solution of the Lagrangian map
  $\bx_f(\bx_i)$ for the example of \eref{eq:numeric}.
\label{fig:vector-field}}
\end{center}
\end{figure}

 However,  a   basic  problem  to   take  into  account   when  solving
\eref{eq:PDF} is that,  if we are given $\rho_i$  and $\rho_f$ then
the normalisation of  the integrals is only  numerically guaranteed to
the  available  precision of  the  computer.   This implies  that  the
numerical solution  ceases to  exist at  the ends  of the  supports of
$\rho_i$ and $\rho_f$.

To have a better control over this difficulty one can solve the
equivalent constraint \eref{eq:const-2}, that in one dimension reads
\begin{equ} \label{eq:field}
  \frac{\d x_f(x_i)}{\d x_i}=\frac{\ri(x_i)}{\rf(x_f(x_i))} ~.
\end{equ}

The problem  of finding the  minimiser map is simply  transformed into
solving  an ODE.  \eref{eq:field}  defines a  vector  field in  the
$(x_i    \,    x_f)$   plane    that    can    be   easily    obtained
numerically\footnote{Here and in what follows, the numerical solutions
  are given in  units of $\kB T$.}.  For any point in  this plane, the
vector field determines the local evolution of \eref{eq:field}. We show
this   in  \fref{fig:vector-field}   for   a   number  of   points
$(x_i \, x_f)$ chosen randomly.

Note that the vector field is  not really defined, outside the central
region  (the   central  lobe  in  \fref{fig:vector-field}), as
the problem has infinite derivatives in the vertical direction. The
same happens at the left and right ends of the central area where the
derivatives are zero and thus, the inverse of the Lagrangian map is
not well defined.

\begin{figure}[!t]
\begin{center}
\includegraphics*[width=0.95\textwidth]{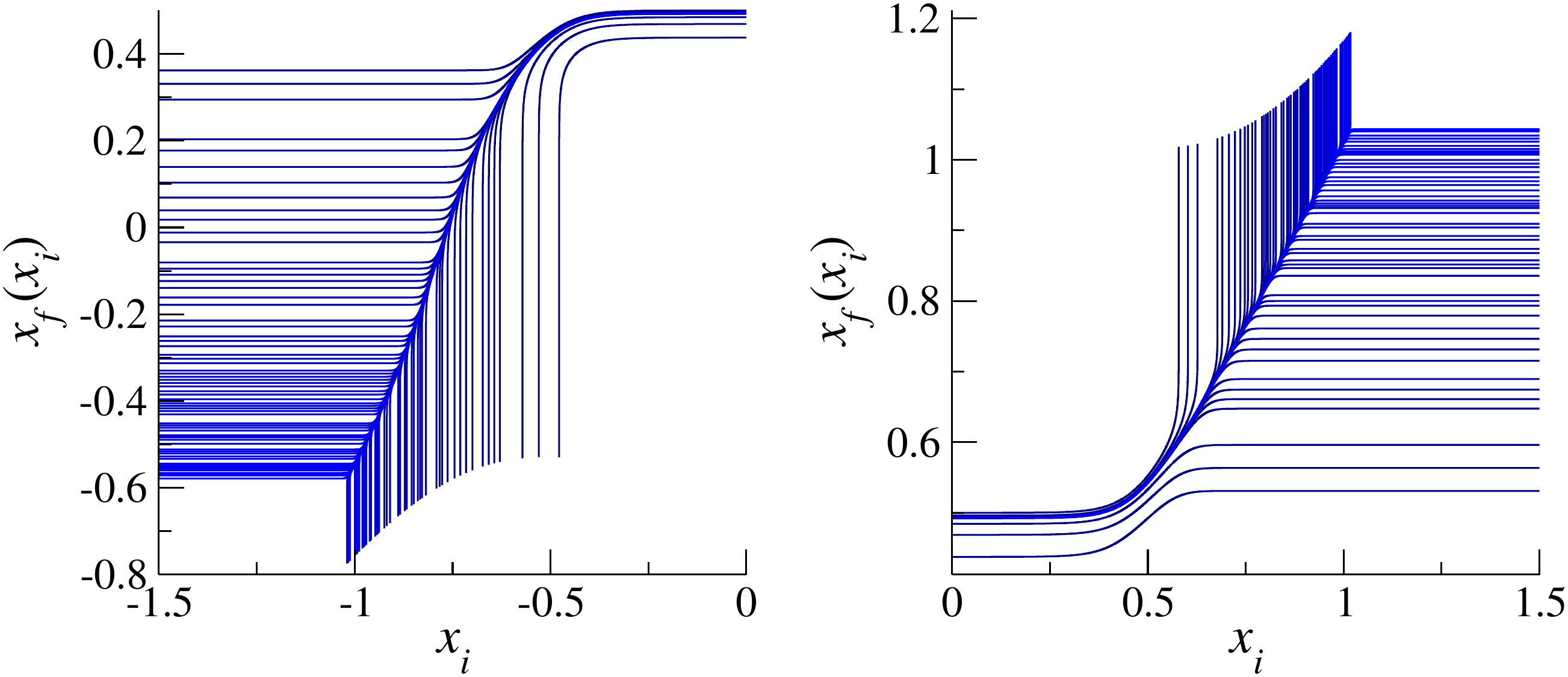}
\caption{The successive improvement of the shooting solution by changing the
  initial  condition until the solution exists over a maximal extent.
\label{fig:bisection}}
\end{center}
\end{figure}

In  view  of \fref{fig:vector-field},  solving  \eref{eq:field}
requires to choose appropriate initial conditions. It is clear that it
is  a good  idea to  start  somewhere in  the center,  at some  height
$y\sim  0.5$  and to  integrate  both  backward  and forward.   Up  to
numerical  precision of  about  $10^{-20}$ we  found that  integrating
backwards from $x_i=0$ fixed, the best numerical initial condition is
$x_f=\,\,\,0.493113178303063601340966142029819$ and when integrating forward\\
$x_f=0.493113178303063601752771313946797$. 

These values represent the numerical possibilities for the example of
\eref{eq:numeric}. They were obtained by shooting as follows: we
start with $x_f=0.5$, which is about the center of
\fref{fig:vector-field}.  We integrate backwards and check for
which $x_i$ the solution ceases to exist, either because the solution
blows up (the graph is vertical), or the solution becomes constant
(the graph is horizontal). We then try another value of $x_f(0)$
(\emph{e.g.}, $x_f=0.46$), measure the divergence value for this new
initial condition, and then solve by bisection for better and better
values of $x_f(0)$ until the value of $x_i$ at which divergence is
observed, cannot be improved any more.

To exemplify our shooting, the approximations through the bisection
procedure are shown in \fref{fig:bisection} for the backward
integration (left panel) and forward integration (right panel).  We
needed about 110 iterations to reach the maximal precision. The speed
of convergence can certainly be improved by \emph{e.g.}, quadratic
interpolation.

\begin{figure}[!t]
\begin{center}
\includegraphics*[width=0.9\textwidth]{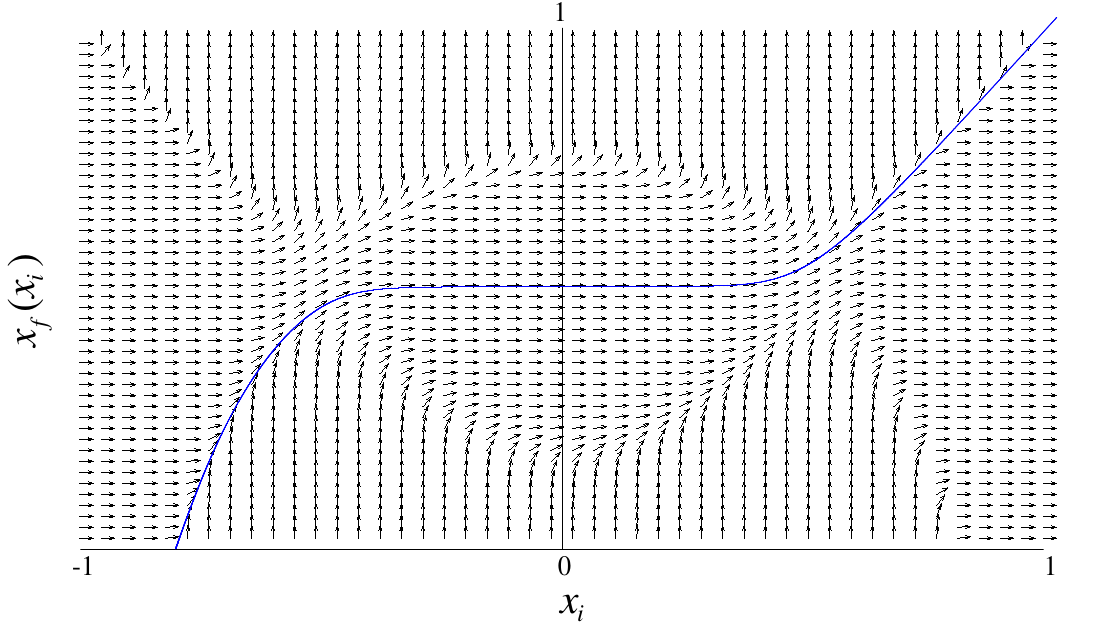}
\caption{The vector field of the flow of the numerical
  convergence, with the minimiser of the Lagrangian map overlaid in blue.
  \label{fig:vector-field-sol}}
\end{center}
\end{figure}

The    solution   we    obtain    in   this    way    is   shown    in
\fref{fig:vector-field-sol}  as  the  blue   curve.  It  yields  a
discrete approximation  of the minimiser of  the continuous Lagrangian
map, that can be improved by increasing the numerical precision. 

Finally, since the current velocity is gradient (see
\eref{eq:grad}), the current velocity is simply the time derivative
of \eref{eq:minL} and the optimal control potential is obtained from
\eref{eq:v} (see Ref.~\cite{Aurell2012} for further details).

\section{Conclusions}

In this paper we have explored the numerical solution of the Monge
problem of optimal transportation, applied to the bit erasure problem
and the Landauer limit explored by Krzysztof in
Ref.~\cite{Aurell2012}.

The densities of \eref{eq:numeric} are appropriate to discuss the
problem of erasing a bit: At the initial time $t=0$, the probability
for a particle to be at the left (around $x=-0.5$), or at the right
(around $x=0.5$), is the same. This corresponds to the bit to be in
state $0$ or $1$ with equal probability, and is well described by a
Gibbs state $\rho_i\propto \exp\left(-\frac{1}{\kB T} V_i(x)\right)$
of \eref{eq:numeric}, with a potential $V_i(x)$ with two symmetric
wells separated by a sufficiently high barrier.  At final time
$t=\tau$, the final Gibbs states $\rho_f$ corresponds to a potential
with only one of the two wells, in our case to a well centered at
$x=0.5$.  This means the final state of the bit is $1$, irrespective
of its initial state.

The particular choice of \eref{eq:numeric} was obtained to reasonably
match those used in the experiment reported in
Ref.~\cite{Berut2012}. The entropy change between $\rho_i$ and
$\rho_f$ is $\Delta S_\mathrm{sys} \approx -0.74312 \kB$, slightly
smaller than $-(\ln 2) \kB$.

To  obtain the  minimal  dissipated heat  during  this transition,  we
reduced  the  optimal mass  transport  in  one  dimensions to  an  ODE
problem, and showed that shooting  allows to find the optimal solution
through  bisection. The  crucial  information to  obtain a  convergent
method was  the knowledge of  the vector  field of the  solution space
(see \fref{fig:vector-field}).

Plugging the solution of the minimiser Lagrangian map into
\eref{eq:cost}, we obtain the minimal cost
$\C_\mathrm{min} = 1.98897268 \kB T$. In conclusion, erasing a bit
through a transformation whose dynamics is described by
\eref{eq:Langevin} evolving in a finite time $\tau$ between the states
of \eref{eq:numeric} can be performed by an average dissipated heat
which is
\begin{equ}
  \langle Q \rangle \ge \left(\frac{1}{\tau} 1.98897268 + 0.74312
  \right) \kB T \ge \kB T \ln 2 \ .
\end{equ}

We consider  the value of the minimal cost that we obtained 
an improvement to the value  $1.996 \kB T$ that Krzysztof obtained
in \cite{Aurell2012}. We wish we could have discussed this with him.

\subsection*{Acknowledgments}
JPE has profited from Swissmap (Fonds National Suisse).  CMM
acknowledges financial support from the Spanish Government grant
PID2021-127795NB-I00 (MCIU/AEI/FEDER, UE), and in part by the
International Centre for Theoretical Sciences (ICTS) for the online
program - "Classical and Quantum Transport Processes: Current State
and Future Directions" (code: ICTS/ctqp2022/1).  We thank E. Hairer
and G. Wanner for very helpful discussions.

\bibliographystyle{plain}
\bibliography{monge}

\end{document}